
\magnification=\magstep1
\overfullrule=0pt

\def\frac#1#2{{#1\over #2}}           
\let\bm=\bf

\font\chapfont=cmbx10 scaled \magstep2

\font\header=cmcsc10
       \font\ninerm=cmr9
       
   \font\nineit=cmti9
\font\dots=cmbx10

\def\nh{\noindent\hang}
\def\eq#1{\eqno\hbox{(#1)}}

\def\r{\vrule height0.6pt depth0pt width1.5cm \  }

\def\fnd#1{{\parindent=10pt\baselineskip=12 true pt\mathsurround=0pt%
    \def\it{\nineit}\footnote{{$^\dagger$}}{\ninerm {#1}}}} %

\outer\def\heading{\bigbreak\bgroup \let \\=\cr \tabskip
      \centering \halign to \hsize\bgroup \header
                 \hfill\ignorespaces## \unskip\hfill\cr}
     \def\endheading{\cr\egroup\egroup \nobreak
                        \medskip\noindent}

\centerline{\chapfont Time Evolution In Macroscopic Systems.}\par
\centerline{\chapfont  I: Equations of Motion} \bigskip
\centerline{\dots W.T. Grandy, Jr.} \bigskip
\centerline{\dots Department of Physics \& Astronomy, University of Wyoming}\smallskip
\centerline{\dots Laramie, Wyoming 82071}
\bigskip

\noindent {\bf Abstract.} Time evolution of macroscopic systems is re-examined primarily through further
analysis and extension of the equation of motion for the density matrix $\rho(t)$. Because $\rho$ contains
both classical and quantum-mechanical probabilities it is necessary to account for changes in both in the
presence of external influences, yet standard treatments tend to neglect the former. A model of
time-dependent classical probabilities is presented to illustrate the required type of extension  to the
conventional time-evolution equation, and it is shown that such an extension is already contained in
the definition of the density matrix.\bigskip

\noindent {\bf 1. Introduction}\medskip

A principal tenet of statistical thermodynamics for over a century has been that the microscopic constituents of
macroscopic systems obey the fundamental dynamical laws of physics, but today there is still no broad
consensus as to exactly how this translates into the time evolution of the macroscopic system itself. At the heart 
of the difficulty is the fact that macroscopic systems require a probabilistic description, so that a primary 
concern must lie with the time development of probabilities themselves, and rarely has this concern been addressed 
from first principles. In a quantum description the problem is further exacerbated by the presence of two 
kinds of probabilities in the density matrix: one intrinsic to the underlying quantum mechanics, and another
pertaining to incomplete information in the context of classical probability theory. The aim of the following 
discussion is to disentangle these two contributions, at least conceptually,  in a manner that leads to unambiguous
equations of motion for
macroscopic systems and at the same time clarifies the foundation of the latter in the microscopic laws.

A classical description of a many-body system begins by introducing an ensemble of like systems along with a phase 
function $f$ exhibiting their distribution in the phase space. We shall find it more convenient to employ a
quantum mechanical description instead, not only because the mathematical expressions are a bit less unwieldy, but 
also because it allows us to focus more readily and naturally on the single system under study. For an {\it isolated}
system\fnd{In many common cases the system can be merely closed, or even open, as long as fluctuations are small; the 
only requirement is that there be no net external forces.}
the standard stages in such a study are the following: (i) construct an initial density matrix $\rho_0$
describing the initial macroscopic state; (ii) let the system evolve under its Hamiltonian $H$, thereby evolving
the density matrix $\rho$ by the deterministic microscopic equation of motion
$$
i\hbar{\dot\rho}=[H,\rho]\,, \eq1
$$
where the superposed dot denotes a total time derivative;
and, (iii) at time $t$ use the time-evolved $\rho(t)$ to predict the expectation value of a system variable $C$ as
$$
\langle C(t)\rangle ={\rm Tr}[\rho(t)C]={\rm Tr}[\rho_0C(t)]\,. \eq2
$$
This last expression illustrates the equivalence of the Schr\"odinger and Heisenberg pictures, for Eq.(1) itself 
is equivalent to a unitary transformation:
$$
\rho(t)=U(t)\rho_0U^\dagger(t)\,, \eq3
$$
where the time-evolution operator $U(t)$ is determined by
$$
\frac{dU}{dt}=HU\,, \eq4
$$
with initial condition $\rho(t_0)=\rho_0$.
For the moment we shall consider only the Schr\"odinger picture in which $\rho(t)$ evolves in time; the density
matrix is very definitely {\it not} a Heisenberg operator. If $\rho$ is stationary, $[H,\rho]=0$, it's a constant 
of the motion; if $\rho$ is a functional only of constants of the motion the macroscopic state is one of 
thermodynamic equilibrium,
corresponding to maximum entropy, and stage (ii) is solved immediately.

It would appear, at least in principle, that proceeding through these three stages must lead to a complete description 
of time-varying processes in a macroscopic system in terms of the underlying microscopic dynamics.
An exact solution of (1) is practicably out of the question for any macroscopic system with a nontrivial Hamiltonian,
of course, so that many approximations have been pursued over the years. Classically the ensemble density of 
$N$-particle systems, $f_N(q,p,t)$, where $\{q,p\}$ represents the collection of all $6N$ coordinates of a system
in phase space, satisfies the Liouville equation: $\partial f_N=\{H,f_N\}$, where the right-hand side is a Poisson 
bracket. Integration over all coordinates but those of a single particle yields the one-particle distribution $f_1$, 
for which an explicit equation of motion at low densities is readily derived, the well-known Boltzmann equation. In 
a quantum mechanical formulation much effort has been devoted to deriving the so-called {\it master} equation for  
a coarse-grained probability distribution over system states (Pauli, 1928). As emphasized by van Kampen (1962),
the solutions describe a single system and there is no need for the notion of ensembles in this approach. Other lines 
of attack include projection operator techniques (Zwanzig, 1961; Mori, 1965), and the notion from the Brussels
school that there may be an intrinsic irreversibility within the microscopic equations themselves. 

All of these
attempts at obtaining macroscopic equations of motion tend to create some sort of irreversibility, but it is difficult
to establish whether it is the real thing or simply an artifact of the approximations. We now demonstrate, however,
that even imagining we could solve (1) exactly leads us into serious difficulties.

There are two simple applications of our 3-stage scenario that lead us directly to some disturbing questions concerning its
general applicability. The first of these involves the response of the system to the presence of a well-defined
external field. The {\it effects} of this field on the system can often be sensibly described by including an additional
term in the Hamiltonian, as in
$$
H_t=H_0-v(t)A\,, \eq5
$$
where $v(t)$ describes the time dependence of the external force and $A$ is a system variable (operator) coupling 
that force to the medium. A formal solution to (1) is given by 
$$
\rho(t)=\rho_0+\frac{i}{\hbar}\int_{t_0}^tU(t,t^\prime)v(t^\prime)\bigl[A,\rho_0\bigr]U^\dagger(t,t^\prime)\,dt^\prime\,,
\eq6
$$
and $t_0$ is the time at which the external force is turned on. The interpretation here is that prior to $t_0$ the system
is in thermal equilibrium, and for $t>t_0$ the density matrix evolves unitarily under the operator $U(t,t^\prime)$ 
determined by (4) with Hamiltonian $H_t$. At any later time the response of the system, described by the departure
of expectation values of other operators $C$ from their equilibrium values, is found by substitution of (6) into
(2) to be 
$$
\langle C(t)\rangle-\langle C\rangle_0 =\int_{t_0}^t\Phi_{AC}(t,t^\prime)v(t^\prime)\,dt^\prime\,, \eq7
$$
where $\langle\ \rangle_0$ denotes an equilibrium expectation value, and
$$
\Phi_{AC}(t,t^\prime)\equiv \frac1{i\hbar}\langle[A,C(t,t^\prime)]\rangle_0 \eq8
$$
is called the {\it dynamical response functional}. The time dependence of $C$ is given by
$C(t,t^\prime)=U^\dagger(t,t^\prime)C(t^\prime)U(t,t^\prime)$, which is effectively now a Heisenberg operator.

Most often $U$ is approximated by $\exp[-i(t-t^\prime)
H_0/\hbar]$, leading to the well-known linear response theory. Although this approximation has been criticized as
inappropriately approximating the microscopic dynamics ({\it e.g.}, van Kampen, 1971), this is really the least of the
problems that arise.
Unquestionably the exact time-evolved $\rho(t)$ will predict the correct value of $\langle C(t)\rangle$ at time $t>t_0$, for
both the quantum mechanics and associated mathematics are impeccable. But, as noted in (3), that time evolution is 
equivalent to a unitary transformation, under which the eigenvalues of $\rho$ are unchanged. These eigenvalues are the
probabilities for the system to be found in any of its possible macrostates under given macroscopic constraints, hence 
there would seem to be no {\it macroscopic} change in the system; but there is such a change, of course, as indicated in (7).
The equivalent classical observation is that the Liouville
equation moves the ensemble distribution around the phase space subject to given constraints, but does not alter those
constraints. A further difficulty is that the von Neumann entropy $S=-k{\rm Tr}(\rho\ln\rho)$, where $k$ is Boltzmann's constant,
is also invariant under unitary
transformation, indicating the absence of irreversible behavior during the process despite the possibility of energy
having been added to the system throughout the time interval. 

A similar but more general application is the common task of heating a pot of water, on an electric stove, say.
To describe this process in complete detail we have to
specify the total system Hamiltonian $H_0$ of the closed system consisting of water, pot,
electric burner, and interactions among them:
 $$
H_0=H_{\rm water}+H_{\rm pot}+H_{\rm burner}+H_{\rm int}\,. \eq9
$$
When the burner is turned on the voltage and current can be enfolded into an
external contribution $H_{\rm ext}(t)$, which leads to a total Hamiltonian
$H_{\rm tot}=H_0+H_{\rm ext}$ for the heating process. With the water initially in thermal equilibrium
with the rest of the system at
temperature $T_i$, we know that the initial density matrix is given by the canonical distribution
$$
\rho(0)=\frac{e^{-H_0/kT_i}}{Z}\,, \qquad Z={\rm Tr}\,e^{-H_0/kT_i}\,. \eq{10}
$$
If the burner is turned on for a period $(0,t)$, then the density matrix for the isolated system at
time $t$ is obtained by unitary transformation, as above.
Upon turning off the switch one expects the system to relax almost immediately to a final equilibrium
state at temperature $T_f$, and the conventional teaching is that  $\rho(t)$
somehow goes over into the final canonical density matrix known to describe thermal equilibrium,
$$
\rho(t)\longrightarrow \rho_f =\frac{e^{-H_0/kT_f}}{Z_f}\,, \qquad Z_f={\rm Tr}\,e^{-H_0/kT_f}\,.
\eq{11}
$$
But this cannot happen: because the eigenvalues of $\rho_f$ and $\rho(t)$ are in general different, the two 
density matrices are incompatible; the eigenvalues of $\rho(t)$ are just those of $\rho_i=\rho(0)$. Indeed, 
as in the previous example, the theoretical entropy of the final state is the same as that of the initial state,
whereas we are certain that the initial and final {\it measured} entropies cannot be the same. Where is the
irreversibility of this process to be found? We shall address that question in Part II of this discussion (Grandy, 
2003; following paper, herein referred to as II);
our task here is to sort out the details of the time evolution process {\it per se} in the presence of
external influences. \bigskip

\noindent {\bf 2. Origin of The Difficulties} \medskip

The source of the above difficulties is that the Hamiltonian governs only the {\it microscopic} behavior of the 
system. While there is little doubt that the $\rho(t)$ evolved under $H$ will make correct predictions of
macroscopic expectation values, it does so by including only the {\it local effects} of the external forces, with
no reference to either the external sources or the macroscopic constraints; and it is the changes in those 
constraints that constitute much of the thermodynamics. Time development by unitary transformation alone affects 
only the basic quantum mechanical probabilities, but not those of classical probability theory that are 
determined by the macroscopic constraints. Similarly, in the classical formulation the ensemble density $f_N$ 
is evolved by Liouville's equation and the meaning of the partial time derivative there is that an observer 
fixed in phase space would see the distribution move by without changing its shape. In either formulation the 
canonical microscopic equations of motion are ultimately responsible for the changing state of the system, to be 
sure, but both the impetus for these changes and the observed effects are macroscopic and must be included as 
such in any realistic analysis.

To explore the origins of these matters more deeply and systematically it will be useful to return to the task 
of stage (i) mentioned earlier, and recall some of the details involved in constructing an initial density matrix, 
or probability distribution.\fnd{For simplicity we shall consider a discrete set of states or alternatives, which 
is equivalent to employing a representation in which $\rho$ is diagonal. The description then appears as
independent of any particular physical application.} We adopt the view that the probability for any one
in a set of $n$ mutually exclusive and exhaustive
alternatives $\{x_i\}$ is contingent on given information $I$, and will be written $P_i=P(x_i|I)$.
As first developed by Gibbs (1902), and elucidated further by Shannon (1948) and Jaynes (1957), $P_i$ is determined 
uniquely by maximizing the {\it information entropy}
$$
S_I=-K\sum_i P_i\ln P_i\,, \qquad K={\rm constant}>0\,, \eq{12}
$$
subject to constraints provided by $I$.
The subscript $I$ emphasizes that this theoretical entropy is defined in the context, and as a part, of
probability theory. This is an important {\it caveat}, for otherwise it is easy to confuse $S_I$ with physical 
or thermodynamic entropy. In fact, merely by making this distinction we see that the invariance of the von Neumann
entropy $S=-k{\rm Tr}(\rho\ln\rho)$ under unitary transformation is not as great a problem as first thought; it,
too, should be considered an $S_I$, and it is only its maximum subject to physical constraints that
corresponds to thermodynamic entropy. Note that $S_I$ has the form of an expectation value, $S_I=-\langle\ln P
\rangle$, which is just the negative of Gibbs' `average index of probability' that he minimized to define the
equilibrium state. 

With the advent of the Shannon-Jaynes insights into construction of prior probabilities based on given
evidence, the reasoning behind the Gibbs algorithm is immediately transparent. Given information in terms of a 
function $f(x)$, such that $x$ can take one, but only one, of the values $\{x_i\}$,
the optimal choice of a probability distribution over $\{x_i\}$
is obtained by maximizing the {\it entropy of the probability distribution} (12) subject to constraints
$$
\sum_{i=1}^n P_i=1\,, \qquad P_i=P(x_i|I)>0\,, \eq{13{\rm a}}
$$
$$
I:\ F\equiv \langle f(x)\rangle=\sum_{i=1}^n P_i f(x_i)\,. \eq{13{\rm b}}
$$
This is the {\it principle of maximum entropy} (PME), and in this form presumes the information to be given in 
the form of an expectation value.

As is well known, the solution to this variational problem is most readily effected by the method of Lagrange
multipliers,  so that the desired probability distribution is given by
$$
P_i=\frac1{Z}e^{-\lambda f(x_i)}\,, \qquad Z(\lambda)=\sum_i e^{-\lambda f(x_i)}\,, \eq{14}
$$
and the normalization factor $Z$ is called the {\it partition function}. Substitution of $P_i$
into (13b) provides the differential equation formally determining the Lagrange multiplier $\lambda$:
$$
F=\langle f\rangle=-\partial\ln Z/\partial\lambda\,, \eq{15}
$$
and the expected value for any other function $g(x)$ is given by $\langle g\rangle=\sum_i P_ig(x_i)$.
It must be stressed that the expectation value on the left-hand side of (13b) is a {\it known
number} $F$ that we have identified in this way so as to incorporate the given information or data
mathematically into a probability distribution. Whether we use one notation or the other will depend on
which feature of these data we wish to emphasize in a particular discussion. This scenario is readily generalized 
to include several pieces of data in terms of several functions $\{f_r\}$, although it is also useful to
note that not all these data need be specified at once. For example, a distribution can be constructed via
the PME based on a datum $\langle f_1\rangle$; subsequently further information may emerge in the form
$\langle f_2\rangle$, and the new distribution is obtained by re-maximizing $S_I$ subject to both pieces of data.
If the new
information contradicts the old there will be no solutions for real $\lambda_2$, and if the new datum is irrelevant
it will simply drop out of the distribution. This procedure provides a method for incorporating new information 
into an updated probability estimate, in the spirit of Bayes' theorem in elementary probability theory.

Maximization of $S_I$ over all probability distributions subject to given constraints transforms the
context of the discussion into one involving the maximum as a function of the data specific to this
application; it is no longer a functional of probabilities, for they have been `integrated out'. To acknowledge this
distinction we shall denote the maximum entropy by $S_{\rm theor}$, and recognize that it is now a function
only of the measured expectation values or constraint variables. That is, $S_{\rm theor}$ is the entropy of the
macrostate, and the impetus provided by information theory is no longer relevant. What remains of the
notion of information is now only to be found on the right-hand side of $P(A|I)$; we are here applying
probability theory, {\it not} information theory. Substitution of (2--4) into (2--2) provides an explicit
expression for the maximum entropy,
$$
S_{\rm theor}=K\ln Z +K\lambda\langle f\rangle\,. \eq{16}
$$

The scenario described by Eq.(14) is precisely that leading to the canonical distribution (10) when the
single piece of data involves the Hamiltonian, or total energy $E(E_i)$, where $\{E_i\}$ is the set
of possible system energies. For constants of the motion, $H$ in this case, the PME provides most of elementary
classical thermodynamics and solves the tasks of stages (i) and (ii) in a single step. Stage (iii), of course, is
very well developed mathematically for equilibrium systems, and we only note here that the expectation value
minimizes the root-mean-square error in estimating $f$.

When the specified functions or operators are not constants of the motion, or they vary in time, then $\rho$ no 
longer commutes with $H$, Eq.(1) must be addressed explicitly, and we return to the conundrums raised above.
Small changes in the problem defined by Eqs.(13) can occur through changes in the set of possible values
$\{f_i\equiv f(x_i)\}$, as well as from changes $\delta p_i$ in the assigned probabilities. A small change in the 
expectation value is then 
$$
\delta\langle f\rangle= \sum_i p_i\delta f_i +\sum_i\delta p_i f_i\,, \eq{17}
$$
and one readily verifies that the corresponding change in the information entropy is
$$
\delta S=S-S_0= \sum_i\delta p_i\ln p_i\,. \eq{18}
$$
The first sum on the right-hand side of (17) is just $\langle\delta f\rangle$, the expected change in $f$, so
we can rewrite that expression as 
$$
\delta\langle f\rangle-\langle\delta f\rangle=\delta Q_f\,, \eq{19}
$$
where $\delta Q_f\equiv\sum_i\delta p_i f_i$. 

Equation (19) can be interpreted as a generalized First Law of thermodynamics, which is suggested by taking
$f=E$, the total system energy. In that case we can interpret $\langle E\rangle$ as the predicted internal energy $U$ and,
since $\delta E_i$ is the work done on the system when it is in state $E_i$, it must be that $\delta W=-\delta\langle 
E\rangle$ is the predicted work done by the system. In this case, then, (19) has the form $\delta U+\delta W=\delta Q$,
and $\delta Q$ is unambiguously identified as heat. The latter is usually thought of as energy transfer through degrees 
of freedom over which we have no control, whereas work takes place through those we do control.
But this is now seen as a special case of a more general rule in
probability theory: a small change in any expectation value consists of a small change in the physical quantity 
("generalized work") and a small change in the probability distribution ("generalized heat"). Just as with ordinary 
applications of the First Law, we see that the three ways to generate changes in any scenario are interconnected, and
specifying any two determines the third. 

The important point for the current discussion is that $\delta Q_f$ is effectively a {\it source term}, and it arises
only from a change in the probability distribution. From (18) it is then clear that any change in the information
entropy can only be induced by a $\delta Q$. Thus, since a unitary transformation corresponding to the time-evolution 
equation (1) leaves $S_I$ invariant, any complete description of the system evolution must contain some explicit 
reference to the sources inducing that evolution. A source serves to change the macroscopic constraints on the 
system, which the microscopic equations alone cannot do, and this can lead to changes in the maximum entropy. In
the case of thermal equilibrium this is
simply thermodynamics at work: a small change in $\langle f\rangle$ induced by a source $\delta Q_f$ results in 
a new macroscopic state corresponding to a re-maximized entropy, a readjustment brought about by the underlying 
particle dynamics, of course.

A deeper issue uncovered here has to do with the nature of probability itself. Many writers subscribe to the view
that objects and systems possess intrinsic probabilities that are actually physical properties like mass or charge.
While most physicists would surely reject such a stance, the idea often seems to lurk in the background of many
probabilistic discussions. One consequence of this viewpoint is that one may be led to believe that a density 
matrix $\rho(t)$ is a physically real quantity that's completely determined by the usual dynamical equations
of motion, rather than representing a state of knowledge about a physical situation. This may work for an isolated 
system for which the only information
available is that encoded in the initial value $\rho(0)$, but we have seen above that this cannot be the case 
when external influences are operative. The probabilities $\langle n|\rho|n\rangle$ can change only if the information 
on which they are based changes. Thus, the resolution of the difficulties discussed above is to be found
through re-examination of time evolution when changes in external constraints are explicitly taken into 
consideration. \bigskip

\noindent {\bf 3. A Time-Dependent Probability Model} \medskip

The question of how to define time-dependent probabilities unambiguously is an old one, but no general theory
seems to have been developed. In the real world any kind of change must have a physical origin, and so most 
expositions tend to focus on, or adapt physical equations of motion in one way or another to describe time 
varying probabilities. But this again risks viewing a probability as a physical object or property, rather than a
representation of a state of knowledge. While quantum mechanical probabilities are governed by microscopic
physical laws, this is not necessarily the case for the macroscopic probabilities of interest here. The point of
this section, then, is to develop a mathematical model that may provide some insight into the type of extension 
of the physical equations that we are seeking.

Our problem is that of defining a time-dependent probability {\it unambiguously}. An understanding that
all probabilities
are conditional on some kind of information leads to the realization that $P(A_i|I)$ can change in time
only if the information $I$ is changing in time, while the propositions $A_i$ are taken as fixed. For example,
if the trajectory $y(t)$ of a particle is changing erratically owing to the presence of an unknown
time-varying field, the allowed values of $y$ do not change, but information on the current position
and some time-dependent effects
of that field might permit construction of probabilities for which values of $y$ might be realized
at some later time $t$.

Armed with this insight the path to extending the PME algorithm in a straightforward manner is clear. We shall 
do this in steps by introducing an abstract probability model that avoids possible confusion with physical 
laws for the time being. Equations (13)-(15) summarized very briefly the maximum-entropy procedure for constructing
a probability distribution based on a single piece of time-independent information.
Unless there is
a definite reason to suppose that the observed value of $\langle f(x)\rangle$ is unvarying, as is the case with
equilibrium statistical mechanics where only constants of the motion are considered, there is little
to persuade us that a subsequent observation will not produce a different value. One might, for example, harbor
such thoughts while rolling a die made from a sugar cube. To generalize our model somewhat let us suppose that
$\langle f(x)\rangle$ is observed at a series of times, and ultimately over a continuous time interval
$[\tau_0,\tau]$. Since there is a piece of data given at every instant in the interval, there is likewise a 
Lagrange multiplier defined at each point as well.
We thus accumulate a body of information that, in the same manner as above, leads to the
maximum-entropy prescription
$$
\eqalign{P_i(\tau_0,\tau) &= Z^{-1}[\lambda] \exp\left\{\int_{\tau_0}^\tau \lambda(t^\prime)f_i(t^\prime)\,
dt^\prime\right\} \cr
Z[\lambda] &=\sum_i \exp\left\{\int_{\tau_0}^\tau \lambda(t^\prime)f_i(t^\prime)\,dt^\prime\right\}\, \cr
\langle f(x,t)\rangle &= \frac{\delta}{\delta\lambda(t)}\ln Z[\lambda(t)]\,, \quad \tau_0\le t\le\tau\,, \cr
\langle g(x)\rangle &= \sum_i P_i(\tau_0,\tau)g_i\,.  \cr} \eq{20}
$$
Thus, $Z$ is now a partition {\it functional}, $\lambda(t)$ a Lagrange-multiplier {\it function} defined
only on the interval $[\tau_0,\tau]$, and this function is determined through functional differentiation.

Although nothing in (20) is to be considered explicitly time dependent, it is true that $\langle f(x,t)
\rangle$, $\lambda(t)$, $f(t)$ vary over the interval $[\tau_0,\tau]$, but we know only that they do so there;
indeed, $\lambda(t)$ is defined {\it only} on that interval. The meaning of $\langle f(x,t)\rangle$ here is that
at each point of the interval we know a definite value of $f(x_i,t)$, and $\lambda(t)$ is determined such
that the mass of the probability distribution resides squarely on these points over that interval. This 
scenario is just a generalization of stage (i) considered earlier and does nothing more than provide an 
initial probability distribution at time $t=\tau$.

Some further features of (20) are worthy of note, beginning with the observation that $\{P_i\}$ goes over into
the uniform distribution as $\tau\to\tau_0$, as it should. If $f\not= f(t)$ it can be removed from the
integrals and Eqs.(20) reduce to the time-independent case.
While physical processes must be causal, it can be shown ({\it e.g.}, Jaynes, 1979) that
logical inferences can propagate either forward or backward in time, as in geology and astrophysics, say. Thus, $\langle g(x)
\rangle$ can not only be predicted for  $t>\tau$, but also retrodicted for $t<\tau_0$. Generally, as $t$
increases beyond $\tau$ the accuracy of predictions made by this distribution can be expected to deteriorate
continually, especially if $f$ continues to vary; only new data can contribute to a better estimate of
$\{P_i\}$ at this point.

In the previous model we thought of the data being collected over a definite time interval, after which we
maximized the entropy.
Having made this first generalization we can see at once the next step. Information gathered in one interval
can certainly be followed by collection in another a short time later, and can continue to be collected in a
series of such intervals, the entropy being re-maximized after each interval. Now let those intervals become
shorter and the intervals between them closer together, so that by an obvious limiting procedure they all
blend into one continuous interval whose upper endpoint is always the present moment.
Thus, there is nothing to prevent us from
imagining a situation in which our information or data are continually changing in time; weather forecasts
come to mind. With experiments performed in a more controlled manner, such information can be
specified in detail and sources turned on and off at will; a common example is the slow heating of that
pot of water.

The leap made here is to imagine re-maximization occurring at every moment, rather
than all at once. There is no fundamental conceptual difference between the two scenarios, however, for in either
case $f(x,t)$ must be known on the set $\{x_i\}$ during the basic information-gathering time interval. Yet,
how do we justify the notion of continual re-maximization? The key point to realize is that there is no 
causal signal involved here, and no physical readjustment to be made. For any imaginable set of constraints 
there is a corresponding unique maximum entropy, just as for a first-order differential equation there is
a unique solution for any given initial condition and it is unnecessary to actually solve the equation to know the
solution exists. When you warm up your half-cup of coffee by pouring more in from the pot, you've just
re-maximized the entropy of the coffee to conform to the new $N$ and $E$.

Without further ado, we now envision continuous data in the form of a time-varying expectation value,
$$
\langle f(x,t)\rangle_t=\sum_i f_i(t)P_i(\tau_0,t)\,, \qquad f_i(t)\equiv f(x_i,t)\,,\quad
 P_i(t)\equiv P(x_i; \tau_0,t)\,. \eq{21}
$$
That is, $f(x_i,t)$ is {\it given} at these points on $[\tau_0,t]$, and is specified to continue so until
further notice. Then, in analogy to (20),
$$
\eqalignno{P_i(t) &= Z^{-1}_t[\lambda]\exp\left\{\int_{\tau_0}^t \lambda(t^\prime)
f_i(t^\prime)\,dt^\prime\right\}\,, &\hbox{(22{\rm a})}\cr
Z_t[\lambda] &= \sum_i \exp\left\{\int_{\tau_0}^t \lambda(t^\prime)
f_i(t^\prime)\,dt^\prime\right\}\,, &\hbox{(22{\rm b})}\cr
\langle f(x,t^\prime)\rangle_{t^\prime} &= \frac{\delta}{\delta\lambda(t^\prime)}\ln Z_{t^\prime}
\bigl[\lambda(t^\prime)\bigr]\,,\qquad \tau_0\le t^\prime\le t\,,&\hbox{(22{\rm c})}\cr
\langle g(x,t)\rangle_t &= \sum_iP_i(t)g_i(t)\,.&\hbox{(22{\rm d})}\cr}
$$
In these expressions the subscript $t$ denotes expectation values computed with $\{P_i(t)\}$, and
we note that the function $g$ need not itself depend explicitly on time. Also, none of these quantities
is necessarily a continuous function of $x$; rather, $x$ simply denotes the sampling space for the discrete set
$\{x_i\}$.

If $\langle f(x,t)\rangle$ is specified to be constant for all time, then $t\to\infty$ and we regain Eqs.(14).
And once again the distribution is uniform at $t=\tau_0$, whereas if the specified time variation is halted at
some time $t=\tau$ Eqs.(20) are regained.

The actual time variation of $P_i(t)$ is described by
$$
\partial_t P_i(t) =\lambda(t) \Delta f_i(t) P_i(t)\,, \eq{23{\rm a}}
$$
where
$$
\Delta f_i(t) \equiv f_i(t)-\langle f(x,t)\rangle_t\,. \eq{23{\rm b}}
$$
We verify that this integrates back into the original $P_i(t)$ by performing a functional integration in
(22), and in doing so obtain the useful alternative expression
$$
Z_t[\lambda]= \exp\left\{\int_{\tau_0}^t\lambda(t^\prime)\langle f(x,t^\prime)\rangle_{t^\prime}\,
dt^\prime\right\}\,. \eq{24}
$$
Equation (23a) has the form of a `master' equation that is often introduced into time-dependent scenarios;
here it is exact.

Direct differentiation in (22d) yields the `equation of motion'
$$
\eqalignno{\partial_t \langle g(x,t)\rangle_t &=\sum_i P_i(t)\bigl[{\dot g}_i(t)-\lambda(t)g_i(t)\Delta f_i(t)
\bigr] \cr
&= \langle {\dot g}(x,t)\rangle_t +\lambda(t) K_{fg}(x,t)\,, &\hbox{(25)}\cr}
$$
where we introduce the equal-time {\it covariance function}
$$
\eqalignno{K_{gf}(x,t) =K_{fg}(x,t) &\equiv \langle g(x,t)f(x,t)\rangle_t -\langle g(x,t)\rangle_t\langle f(x,t)
\rangle_t \cr
&=\frac{\delta}{\delta\lambda(t)}\langle g(x,t)\rangle_t\,. &\hbox{(26)}\cr}
$$
Note that one can choose $g=f$; otherwise, $g$ need not depend explicitly on $t$, in which case the first
term on the right-hand side of (25) vanishes. While of formal interest, these equations of motion are somewhat
redundant in view of (22d); in physical applications, however, they lead to several important insights.

Because $\{P_i\}$ is now time dependent the information-theoretic entropy analogous to (16), and which was
maximized continuously to obtain Eqs.(22), also depends on the time. There is no implied relation to the
thermodynamic entropy, of course, so we introduce yet another entropic notation and write this functional
as
$$
H_t[P(y)]=-K\sum_{i=1}^n P_i(t)\ln P_i(t)\,, \qquad K>0\,. \eq{27}
$$
Upon maximization this depends on the initial data only, so is a functional of $\langle f(x,t)\rangle_t$.
Substitution from (22) then yields
$$
\eqalignno{H_t[\langle f(x,t)\rangle] &= \ln Z_t[\lambda] -\int_{\tau_0}^t \lambda(t^\prime)\langle f(x,t^\prime)
\rangle_t\,dt^\prime \cr
&= \int_{\tau_0}^t \lambda(t^\prime)\langle f(x,t^\prime)\rangle_{t^\prime}\,dt^\prime -\int_{\tau_0}^t
\lambda(t^\prime)\langle f(x,t^\prime)\rangle_t\,dt^\prime\,, &\hbox{(28)}\cr}
$$
the integrands differing only in the subscripts.
Note that a functional differentiation yields an alternative expression for $\lambda(t)$,
$$
\lambda(t)=\frac{\delta H_t}{\delta\langle f(x,t)\rangle_t}\,, \eq{29}
$$
and a time derivative provides a rate of `entropy production'
$$
\partial_t H_t=-\lambda(t)\int_{\tau_0}^t \lambda(t^\prime)K_{ff}(t,t^\prime)\,dt^\prime\,. \eq{30}
$$

These equations are remarkably similar to many of those found in writings on irreversible thermodynamics 
({\it e.g.}, de Groot and Mazur, 1962). Though no application to physical models is made here,
one recognizes the analogs of fluxes and forces, and Onsager-like reciprocity is immediately evident in
(26). But no linear approximations are made in this model, so the current scenario is considerably more general.
It must be emphasized once again, however, that the time dependencies derived above are based entirely on the
supposition of information supplied in the form of {\it specified} time-varying expectation values or source
functions; we actually know how $f(x,t)$ varies in time, allowing us to predict the variation of $g(x,t)$.
A possible general application of these considerations might be made to driven noise, in which the noise 
amplitude varies in a known way. \bigskip

\noindent {\bf 4. The Physical Problem} \medskip

Precisely how to adapt the preceding probability model to macroscopic systems will be taken up in Part II of
this discussion, while here we shall complete the study of Eq.(1) and the fundamental time-evolution equation. 
If we believe that only an external source can produce changing macroscopic constraints and time-varying information 
$I(t)$, then $\rho(t)$ must evolve in a manner over and above that determined by the Hamiltonian alone. In fact,
such an additional evolution is already implied by the density-matrix formalism, as we now demonstrate.

In the equation of motion (1) for the density matrix we noted that the superposed dot represented a {\it total} time
derivative. But in many works the equation is commonly written as
$$
i\hbar\partial_t\rho =[H,\rho]\,, \eq{31}
$$
where $H$ may be time dependent. The standard argument is that the derivative in (31)
is indeed a partial derivative because this expression is derived directly from the
Schr\"odinger equation, which contains a partial time derivative, although it makes no
difference in (31) since $\rho$ depends only on $t$. This comment would not be notable
were it not for an additional interpretation made in most writings on
statistical mechanics, where $\rho$ describes the entire macroscopic system.

Equation (31) is often compared with the Heisenberg equation of motion for an operator $F(t)$ in the
Heisenberg picture
$$
\frac{dF}{dt}= \frac1{i\hbar} [F,H]+\partial_t F\,, \eq{32}
$$
whereupon it is concluded from analogy with Liouville's theorem that $d\rho/dt=0$ and (31) is just the 
quantum mechanical version of
Liouville's equation in classical statistical mechanics. But there is nothing in quantum mechanics that requires
this conclusion, for $\rho(t)$ is not a Heisenberg operator; it is basically a
projection operator constructed from state vectors, and in any event (31) refers to the Schr\"odinger picture.  
Heisenberg operators are analogous to functions of phase
in classical mechanics, $\rho$ is not. We shall argue here that
the derivative in (31) should be considered a total time derivative, as asserted earlier for (1); this
follows from a careful derivation of that equation.

A density matrix represents a partial state of knowledge of a system. Based on
that information we conclude that with probability $w_1$ the system may be
in a pure state $\psi_1$, or in state $\psi_2$ with probability $w_2$, etc.
Although the various alternatives $\psi_i$ are not necessarily mutually
orthogonal, they can be expanded in terms of a complete orthonormal set
$\{u_k\}$: 
$$
\psi_i({\bm r},t)=\sum_k a_{ik}(t)u_k({\bm r})\,, \eq{33}
$$
such that $\langle u_k|u_j\rangle=\delta_{kj}$. The quantum mechanical expectation value of a
Hermitian operator $F$ in state $\psi_i$ is
$$
\langle F\rangle_i\equiv\langle\psi_i|F|\psi_i\rangle=\sum_{k,n}a^{\rm \hphantom{*}}_{ki}a^*_{ni}
\langle u_n|F|u_k\rangle\,. \eq{34}
$$
The expected value of $F$ over all possibilities (in the sense of classical probability theory)
is then
$$
\langle F\rangle=\sum_i w_i\langle F\rangle_i\,, \qquad \sum_i w_i=1\,. \eq{35}
$$
This last expression can be written more compactly (and generally) in matrix form as
$$
\langle F\rangle={\rm Tr}(\rho F)\,, \eq{36}
$$
where the {\it density matrix} $\rho$ is defined in terms of its matrix elements:
$$
\rho_{kn}\equiv \sum_i a^{\rm \hphantom{*}}_{ki}a^*_{ni} w_i\,. \eq{37}
$$
This expression is equivalent to writing $\rho$ as a weighted sum of projection operators 
onto the states $\psi_i$: $\rho=\sum_i w_i|\psi_i\rangle\langle\psi_i|$.

To find an equation of motion for $\rho$ we recall that each $\psi_i$ must satisfy the
Schr\"odinger equation $i\hbar\partial_t\psi_i=H\psi_i$, and from (33) we find that this
implies the equations of motion
$$
i\hbar {\dot a}_{ij}=\sum_k a_{ik} H_{jk}\,, \qquad H_{jk}\equiv \langle u_j|H|u_k\rangle\,.
\eq{38}
$$
The superposed dot here denotes a total time derivative, for $a_{ij}$ describes a particular
state and depends only on the
time. One thus derives an equation of motion for $\rho$ by direct differentiation in (37),
but this requires some prefatory comment.

Usually the weights $w_i$ are taken to be constants, determined by some means outside the
quantum theory itself. In fact, they are probabilities and can be determined in principle
from the PME under constraints representing information characterizing the state
of the system. As noted earlier, however, if that information is
changing in time, as with a nonequilibrium state, then the probabilities will also be
time dependent. Hence, quite generally one should consider $w_i=w_i(t)$; if such time
dependence is absent we recover the usual situation.

An equation of motion for $\rho$ is now found in a straightforward manner,
with the help of (37) and (38), by computing its total time variation:
$$
i\hbar{\dot\rho}_{kn}=\sum_q\big(H_{kq}\rho_{qn}-H_{nq}\rho_{kq}\bigr)+i\hbar\sum_i{\dot w}_i
a^{\rm \hphantom{*}}_{ki}a^*_{ni}\,, \eq{39}
$$
or in operator notation
$$
i\hbar{\dot\rho}=[H,\rho] +i\hbar\partial_t\rho\,. \eq{40}
$$
The term $i\hbar\partial_t\rho$ is meant to convey only the time variation of the $w_i$.

Comparison of (40) with (32)confirms that the former is {\it not} a Heisenberg equation
of motion --- the commutators have opposite signs. Indeed, in the Heisenberg picture the only time
variation in the density matrix is given by $\partial_t\rho$, which arises qualitatively in the same general 
manner as $P_i(t)$ in the preceding probability model.
If, in fact, the probabilities $w_i$ are constant,
as in equilibrium states, then (40) verifies (31) but with a total time derivative.
Otherwise, (40) is the general equation of motion for the density matrix, such that the first term
on the right-hand side describes the usual unitary time development of an isolated system. The
presence of external sources, however, can lead to an explicit time dependence as represented
by the second term, and thus the evolution is not completely unitary; classically, Liouville's theorem is 
not applicable.
An additional source term of this kind also appears in the work of Zubarev, {\it et al} (1996), and
in Kubo, {\it et al} (1985), but of a considerably different origin and unrelated to the basic probabilities.
\bigskip \goodbreak

\noindent {\bf 5. Summary} \medskip

Equation (40) is the sought after extension to
macroscopic systems of the equation of motion for the density matrix . The difference between this equation 
and the canonical version (1) is evident; it is the differences in their solutions that is most important. 
No matter what the solution to (1), it is always equivalent to a unitary transformation of the initial density 
matrix, and hence incapable of describing an irreversible process completely; some approximations of that equation, 
however, may exhibit various aspects of irreversible behavior. The $\rho(t)$ evolved by (1) in conjunction with 
the total Hamiltonian is certainly a correct result of quantum mechanics, but from a macroscopic viewpoint
it is incomplete; it contains no new macroscopic information about the processes taking place. It can predict 
changes in the macroscopic constraints, yet is not itself affected subsequently by those changes; it reflects changes in the 
quantum mechanical probabilities but not in the $w_i$ in (37), which are determined by the external constraints. 
An example illustrating these points is given in II. 

To elaborate further on these differences, let us recall Boltzmann's enormously insightful relation between the 
maximum entropy and phase space volumes (or manifolds in a Hilbert space). In the form articulated by Planck 
this is 
$$
S_{\rm B}=k\ln W\,, \eq{41}
$$
where $W$ is a measure of the set of microscopic states compatible with the macroscipic constraints on the 
system; it is a multiplicity factor. We emphasize that this expression is to be a representation of the 
{\it maximum} of the information entropy and, as Boltzmann himself observed, it is not restricted to equilibrium 
states. Although it sometimes is stated that (41) constitutes a proper definition of time-dependent entropy for
nonequilibrium states ({\it e.g}, Lebowitz, 1999), there is no theoretical or mathematical basis for this 
assertion. Rather, Boltzmann's formulation --- which can actually be expressed as a theorem ({\it e.g.},
Grandy, 1980) --- provides a deep physical explanation of what is achieved by maximizing the information entropy 
in the manner of Gibbs; namely, $S_{\rm B}$ characterizes that huge set of microstates compatible with the 
macroscopic constraints, each microstate contributing to $W$ having probability roughly equal to $W^{-1}$.
In addition, (41) also illustrates through Liouville's theorem on conservation of phase
volume why the maximum entropy itself remains unchanged under canonical (or unitary) transformation; that is, no 
macroscopic information is either gained or lost during evolution under (1). Yet, (41) {\it must} change under 
the action of external forces; but how? 

Let us return to the scenario of heating a pot of water on an electric burner, where initially the entire system 
is in equilibrium with multiplicity $W_i$. As energy $\Delta Q$ is added to the water its temperature rises and 
the number of macroscopic configurations corresponding to the new constraints increases enormously, until at 
some point the plug is pulled and the entire system relaxes to a final equilibrium state with multiplicity 
$W_f\gg W_i$. As a consequence, the entropy increases to $S_{\rm B}(final) > S_{\rm B}(initial)$, and this is 
completely equivalent to re-maximizing the information entropy subject to the constraint $E_f=E_i+\Delta Q$. 
Note, however, that one can imagine carrying out such a re-maximization at any time during the process, for 
that maximization is also equivalent to acknowledging the existence of a definite phase volume of compatible
microscopic states at any instant. 

Thus, the multiplicity factor $W$ increases to its final value owing to a change in the macroscopic constraint 
provided by the total energy. In turn, this can come about only by an evolution of the weights $w_i$ in (39).
Because the only time variation in $\rho(t)$ in the Heisenberg picture is that of these weights,  we suspect
there may be more direct ways to determine the appropriate density matrix than by trying
to solve an incredibly complex differential equation. After all, from a macroscopic standpoint we need only know
$\Delta Q$ and the heat capacity of the water to predict its final temperature and energy.
Explicit construction of $\rho(t)$ and $S(t)$ appropriate for describing nonequilibrium
phenomena in these systems is carried out in Part II (following paper).

\bigskip\medskip
\centerline{\bf REFERENCES}
 \bigskip

\nh de Groot, S.R. and P. Mazur (1962), {\it Nonequilibrium Thermodynamics}, North-Holland, Amsterdam.

\nh Gibbs, J.W. (1902), {\it Elementary Principles in Statistical
Mechanics}, Yale University Press, New Haven, Conn.

\nh Grandy, W.T. (1980), "Principle of Maximum Entropy and Irreversible Processes," Phys. Repts. {\bf 62}, 175.

\nh\r (2003), "Time Evolution in Macroscopic Systems. II: The Entropy," following paper.

\nh Jaynes, E.T. (1957), ``Information Theory and Statistical Mechanics,''
Phys. Rev. {\bf 106}, 620.

\nh Kubo, R., M. Toda, and N. Hashitsume (1985), {\it Statistical Physics II}, Springer-Verlag, Berlin.

\nh van Kampen, N.G. (1962), ``Fundamental Problems in Statistical Mechanics of Irreversible Processes,"
in {\it Fundamental Problems in Statistical Mechanics}, E.G.D. Cohen (ed.), North-Holland, Amsterdam; p.173.

\nh\r (1971), ``The case against linear response theory," Physica Norvegica {\bf 5}, 279.

\nh Lebowitz, J.L. (1999), ``Statistical mechanics: A selective review of two central issues," Rev. Mod. Phys.
{\bf 71}, S346.

\nh Mori, H. (1965), ``Transport, Collective Motion, and Brownian Motion," Prog. Theor. Phys, {\bf 33}, 423.

\nh Pauli, W. (1928), ``\"Uber das H-Theorem vom Anwaschen der Entropie vom Standpunkt der neuen Quantenmechanik," 
in {\it Probleme der modernen Physik}, Arnold Sommerfeld zum 60. Geburtstage gewidmet vonsiner Sch\"ulern, 
Verlag, Leipzig.

\nh Shannon, C. (1948), ``Mathematical Theory of Communication," Bell System Tech. J. {\bf 27},379, 623.

\nh Zubarev, D., V. Morozov, and G. R\"opke (1996), {\it Statistical Mechanics of Nonequilibrium Processes.
Volume 1: Basic Concepts, Kinetic Theory}, Akadamie Verlag, Berlin.

\nh Zwanzig, R. (1961), ``Statistical Mechanics of Irreversibility," in {\it Lectures in Theoretical Physics, Vol.III},
W.E. Brittin, B.W. Downs, and J. Downs (eds.), Interscience, New York.

\bye